\documentclass[12pt]{article}
\usepackage{latexsym}
\usepackage{epsfig,amssymb,euscript}
\usepackage{amsmath}
\usepackage{mathrsfs}
\usepackage[margin=30pt,font=small,labelfont=bf]{caption}
\numberwithin{equation}{section}  
\newsavebox{\ns}
\newsavebox{\dbrane}

\def\be{\begin{equation}}
\def\ee{\end{equation}}
\def\bea{\begin{eqnarray}}
\def\eea{\end{eqnarray}}

\renewcommand{\theequation}{\arabic{section}.\arabic{equation}}
\def\theequation{\thesection.\arabic{equation}}

\def\Dslash{\,\,{\raise.15ex\hbox{/}\mkern-12mu D}}
\def\Dbarslash{\,\,{\raise.15ex\hbox{/}\mkern-12mu {\bar D}}}
\def\delslash{\,\,{\raise.15ex\hbox{/}\mkern-9mu \partial}}
\def\delbarslash{\,\,{\raise.15ex\hbox{/}\mkern-9mu {\bar\partial}}}
\def\pslash{\,\,{\raise.15ex\hbox{/}\mkern-9mu p}}
\def\calDslash{\,\,{\raise.15ex\hbox{/}\mkern-12mu {\cal D}}}

\newcommand{\nn}{\nonumber \\}

\begin{document}

\makeatletter
\renewcommand{\theequation}{\thesection.\arabic{equation}}
\@addtoreset{equation}{section}
\makeatother

\baselineskip 18pt

\begin{titlepage}

\vfill

\begin{flushright}
Imperial/TP/2010/JS/01\\
\end{flushright}

\vfill

\begin{center}
   \baselineskip=16pt
   {\Large\bf  A gravity derivation of the Tisza-Landau Model in AdS/CFT}
  \vskip 1.5cm
      Julian Sonner$^{1),2)}$  and Benjamin Withers$^{1)}$\\
   \vskip .6cm
        \begin{small}
      {}$^{1)}$\textit{Theoretical Physics Group, Blackett Laboratory, \\
        Imperial College, Prince Consort Rd, London SW7 2AZ, U.K.}
        \vskip2em
        {}$^{2)}$\textit{Trinity College, University of Cambridge, \\
        Cambridge, CB2 1TQ, U.K.}
        \end{small}\\*[.6cm]
   \end{center}

\vfill

\begin{center}
\textbf{Abstract}
\end{center}

\begin{quote}
We derive the fully backreacted bulk solution dual to a boundary superfluid with finite supercurrent density in AdS/CFT. The non-linear boundary hydrodynamical description of this solution is shown to be governed by a relativistic version of the Tisza-Landau two-fluid model to non-dissipative order. As previously noted, the phase transition can be both first order and second order, but in the strongly-backreacted regime at low charge $q$ we find that the transition remains second order for all allowed fractions of superfluid density.

\end{quote}

\vfill

\end{titlepage}
\setcounter{equation}{0}

\section{Introduction}

A holographic superfluid \cite{Hartnoll:2008vx,Hartnoll:2008kx} is a charged, asymptotically AdS black hole, which below a certain temperature $T_c$ spontaneously develops a scalar condensate. The relevant instability of charged scalars in AdS-black-hole backgrounds was first pointed out in \cite{Gubser:2008px}.  In many contexts one can also use such systems to explore the physics of holographic superconductivity. In this work we focus on the superfluid aspects by deriving the boundary superfluid hydrodynamical equations that govern them.

According to the AdS/CFT correspondence \cite{Maldacena:1997re}, asymptotically AdS black holes are dual to thermal states of a large-$N$ boundary conformal field theory \cite{Witten:1998zw}. If we restrict ourselves to considering low-frequency fluctuations of this field theory with large wavelengths, we expect an effective hydrodynamic description to be valid \cite{Policastro:2001yc}. Said differently, the hydrodynamical description is applicable for configurations of the field theory which are locally in thermodynamic equilibrium and which vary from local patch to local patch such that all gradients remain small. It was shown in \cite{Bhattacharyya:2008jc} that such a field-theory gradient expansion can be implemented in AdS/CFT as a perturbative scheme of solving Einstein's equations in the universal subsector of  gravity with a cosmological constant. Charged branes, and thus chemical potentials for conserved global currents in the boundary theory, were introduced by \cite{Banerjee:2008th, Erdmenger:2008rm}. The setup was subsequently generalised in \cite{Hansen:2008tq} to include effects a bulk $U(1)$ field inducing boundary magnetic fields.

A qualitatively rather different picture should arise for systems that allow for the breaking of a global $U(1)$ symmetry thus giving rise to superfluidity. The hydrodynamic limit of such a system is given by the Tisza-Landau \cite{Landau:1941,Tisza:1947zz} two-fluid model, which describes the superfluid as a two-component mixture of a normal fluid with density $\rho_n$ and velocity $u_\mu$ and a superfluid component with density $\rho_s$ and flow velocity $v_\mu$.

 In this paper we find solutions of $4d$ gravity\footnote{The spacetime coordinates $\{ x^M\}$ split into $\{ r,x^\mu \}$, where $r$ is the holographic direction and we refer to $\{x^\mu\}$ as the field-theory or boundary coordinates.} which we argue to be dual to holographic superfluids with finite supercurrent density and we show that their hydrodynamical description is captured precisely by a relativistic version of the Tisza-Landau two-fluid model. We do this by constructing asymptotically AdS solutions that break the bulk $U(1)$ symmetry with a charged condensate at finite chemical potential. The use of charged AdS black holes to study field theory at finite chemical potential was considered early on in \cite{Chamblin:1999tk}, albeit in the context of global AdS. Our solutions are gauge-equivalent to situations where the phase of the bulk scalar, given the interpretation of the field-theory Goldstone mode, has non-trivial dependence on the field theory directions. To non-dissipative order in the fluid expansion this dependence is linear so that the Goldstone mode has a constant gradient.
 
 Some properties of holographic superfluids, both with s-wave and p-wave condensates, such as its various sound modes etc., were explored in \cite{Herzog:2008he,Basu:2008st,Herzog:2009md,Karch:2008fa,Yarom:2009uq,Amado:2009ts,Herzog:2009ci,Arean:2010xd}. Our treatment is complementary to theirs in two ways: Firstly we improve on the state of the art by including the backreaction consistently, that is we solve the full set of coupled Einstein-Maxwell-scalar equations. Secondly, our approach, akin to that of \cite{Bhattacharyya:2008jc} for the normal fluid case, derives the non-linear boundary fluid dynamics, including the `Josephson equation' from bulk gravity, rather than taking them as a given and then fitting transport coefficients to the postulated model.

The paper is organised as follows. Immediately following this paragraph we give a brief overview of the analytical results of this paper. In section 2 we review and extend the by now well-known holographic superconductor. In section 3 we describe the gravity setup corresponding to introducing a non-zero supercurrent on the boundary, perform its holographic renormalisation and define the appropriate statistical grand partition function. In this section we also derive the boundary stress tensor and $U(1)$ current. Section 4 contains our numerical method and results and section 5 gives a discussion of the results and some open issues. Appendix A contains the bulk equations in  their full glory.

\subsection{Summary of results}
Here we summarise the findings of the rest of the paper. We emphasise that the equations of fluid dynamics, as well as the form of the constitutive relations are exact and derived analytically. The hydrodynamical equations following from our bulk analysis to non-dissipative order can be summarised as follows
\bea
T_{\mu\nu} &=& \left( \epsilon + P \right)u_\mu u_\nu + P \eta_{\mu\nu} + \mu \rho_s v_\mu v_\nu \nn
J_\mu	&=& \rho_n u_\mu + \rho_s v_\mu\,, \nn
\partial_\mu J^\mu &=& 0\nn
\partial_\mu T^{\mu\nu} &=& 0
\eea
where the total charge density is $\rho = \rho_n + \rho_s$ and we have set a possible external applied field that can appear on the right-hand side of the stress-tensor conservation equation, to zero. The quantity $u_\mu$ has the interpretation of normal fluid velocity, whereas $v_\mu$ is the superfluid component. These satisfy the constraint
\be
u^\mu v_\mu = -1
\ee
also known as a Josephson equation. The rest of this paper explains the bulk origin of these equations and numerically constructs the dual solutions. We also use our non-linear solutions to study certain physical properties, such as the order of the phase transition, numerically.


\section{Isotropic holographic superfluids}


In this section we review and (slightly) extend the work of \cite{Hartnoll:2008vx,Hartnoll:2008kx} on the equilibrium properties of holographic superfluids. This section also serves to introduce our notation and conventions: rather than using standard Schwarzschild-like coordinates, we use ingoing Eddington-Finkelstein coordinates and introduce a boost $u_\mu$ along the direction of the (flat) horizon. The class of solutions we are constructing in this section are thus generalisations of boosted black branes allowing for a non-trivial profile of a complex scalar field.

 A simple action that is sufficiently general for our aims is given by
\be\label{eq:bulk_action}
S_{\rm bulk} =\int \sqrt{-g}  \left[  \frac{1}{16\pi G}\left(R + \frac{6}{\ell^2} \right) - \frac{1}{4}F_{MN}F^{MN}  - V\left( |\psi|\right)  - |D\psi|^2\right]d^{4}x\,,
\ee
supplemented by the Gibbons-Hawking-York boundary term.
Here $\psi$ is a complex scalar with charge $q$ under the $U(1)$ gauge field
\be
D\psi = d\psi - i q A \psi
\ee
and the potential $V(|\psi|)$ is chosen to consist of just a mass term
\be
V(|\psi|) = -\frac{2}{\ell^2}|\psi|^2\,,
\ee
where $\ell$ is the AdS length.
 Different choices of the potential  \cite{Horowitz:2008bn,Franco:2009yz}  are possible. The work of \cite{Herzog:2009md} used first-order perturbation theory to study the effects of a $|\psi|^4$ term on the various superfluid sound modes. Embedding in M-theory \cite{Gauntlett:2009dn,Gauntlett:2009bh} or string theory \cite{Gubser:2009qm} dictates a specific form of the scalar potential and it would be interesting to investigate their hydrodynamics.
 
 As we are interested in black holes with flat horizon sections, we propose an ansatz of the form
\bea
ds_{\rm static}^2 &=& -2 h u_\mu dx^\mu dr - \tfrac{r^2}{\ell^2} f u_\mu u_\nu dx^\mu dx^\nu + \tfrac{r^2}{\ell^2} \Delta_{\mu\nu} dx^\mu dx^\nu\nn
A &=& -\phi u_\mu dx^\mu - \frac{\ell^2\phi h}{r^2 f}dr \nn
\psi &=& |\psi|\,  e^{iq\alpha} := \xi\,  e^{iq\alpha} \,.
\eea
We have chosen to write everything covariantly in the boundary directions $\{  x^\mu\}$ by introducing the boost velocity $u^\mu$ which satisfies $ \gamma_{\mu\nu}^\infty u^\mu u^\nu=-1$. The quantity $\gamma^\infty_{\mu\nu} := \lim_{r\rightarrow\infty}\frac{\ell^2}{r^2}\gamma_{\mu\nu}$ is the metric on the boundary
\be
ds^2_\infty = \gamma^\infty_{\mu\nu} dx^\mu dx^\nu \,.
\ee
Finally, we have the spatial projector
\be\label{eq:projector}
\Delta_{\mu\nu} = \gamma^\infty_{\mu\nu} + u_\mu u_\nu
\ee
At this point this is nothing more than a boosted version (by $u^\mu$) of the holographic superconductor solution of \cite{Hartnoll:2008kx} written in Eddington-Finkelstein coordinates.

If we rescale $A_\mu \rightarrow \frac{A_\mu}{q}$ and $\psi \rightarrow \frac{\psi}{q}$, the rescaled Maxwell-scalar sector is down by a factor of $q^{-2}$ with respect to the Einstein-Hilbert action. Taking the limit $q\rightarrow\infty$ thus eliminates any backreaction on the metric. Gravity backreaction is physically important not only for low charge $q$ but also at low temperatures. In this paper we are able to work away from the probe limit. For completeness we demonstrate that our results do indeed go over into the various perturbative results known in the literature in the appropriate limits.
\subsection{Exact black brane solution}
The normal phase of the superfluid which has vanishing order parameter on the boundary is given by the standard charged RNAdS black brane
\bea
f &=& h_\infty^2 - \frac{8 \pi G\ell^2 \varepsilon}{r^3} + \frac{4 \pi G h_\infty^2 \ell^6 \  \rho^2}{r^4}\,,\nn
\phi &=& h_\infty\left(\mu - \frac{\rho\ell^2}{r}\right)\,,\quad h = h_\infty \,,\quad \psi = 0\,.
\eea
The metric on the boundary is now
\be
ds^2_\infty = -h_\infty^2 dv^2 + \delta_{ij}dx^i dx^j\,.
\ee
In order for $A$ to be a well-defined one form at the horizon ($r=r_+$), we must set
\be
\rho = \frac{r_+ \mu}{\ell^2}\,.
\ee
 This can be seen, {\it e.g.} by going to Kruskal coordinates. In this gauge $\mu$ can be identified as the chemical potential of the boundary theory. Also $\rho$ is the charge density of the system
\be
\int *F = \rho \  {\rm vol}_2
\ee
with ${\rm vol}_2$ the divergent two-volume of the flat spatial sections of the geometry.
The Euclidean solution can be thought of as a thermodynamic ensemble at temperature
\be
T = \frac{r_+^2 }{4 \pi \ell^2 h_\infty}\frac{f'(r_+)}{ h(r_+)}\,.
\ee
We will soon generalise the metric ansatz, but the expression for temperature will remain unaltered.
\subsection{Holographic stress-energy and current }
It is well known that the naively defined boundary stress tensor in asymptotically AdS space needs to be renormalised by introducing suitable counterterms \cite{Balasubramanian:1999re}. For our system (\ref{eq:bulk_action}), the desired stress tensor and current follow from varying the action
\be
S_{\rm ren} = S_{\rm bulk} + \frac{1}{8 \pi G} \int_{\partial \Sigma} \sqrt{-\gamma}\left( {\cal K} + \frac{2}{\ell}\right) \, d^3 x  + \int_{\partial \Sigma} \sqrt{-\gamma} \, \left(\frac{1}{\ell} |\psi|^2 \right)d^3x\,,
\ee
where $\gamma_{\mu\nu}$ is the induced metric on the boundary. In addition to the standard gravitational terms there is an additional counterterm for the scalar field. This choice is valid provided that one fixes the coefficient of the leading $r^{-1}$ behaviour of $\psi$ to zero. We restrict to this boundary condition from now on. Alternatively, we could have fixed the coefficient of the $r^{-2}$ behaviour by adding a different counterterm for the scalar field $\psi$. The mass of the field $\psi$ places it in the range where both behaviours can be thought of as vevs. For a more extensive discussion, involving both possible choices of boundary conditions, see {\it e.g.} \cite{Hartnoll:2008kx,Gauntlett:2009bh}.
 In order to evaluate the stress tensor and current, we need to determine the asymptotic expansions of the fields. The near-boundary expansion of the equations in appendix \ref{app:bulkequations} yields
 \bea\label{eq:expansion_one}
 h &=& h_\infty\left(1 - \frac{4 \pi G \ell^2 \xi_1^2}{r^2}+ \cdots \right) \nn
 f &=& h_\infty^2 - \frac{8 \pi G \ell^2 \varepsilon}{ r^3}+ \cdots  \nn
 \phi  &=&h_\infty\left(\mu - \frac{\rho \ell^2}{r} + \cdots  \right)\nn
 \xi &=&  \frac{\xi_1}{r} + \frac{\xi_2}{r^2} + \cdots
 \eea
Furthermore, we can set $h_{\infty}$ to unity by invoking a scaling symmetry and we will do so wherever convenient. However, it is important to keep in mind where it occurs formally when we compute boundary variations later on and variations of $h_\infty$ translate into variations of temperature.

It now follows that the current and stress tensor
\bea\label{eq:StressAndCurrent}
T_{\mu\nu} 	&=& \frac{2}{\sqrt{-\gamma}} \frac{\delta S_{\rm ren}}{\delta \gamma^{\mu\nu}} = \frac{1}{8 \pi G} \left[ {\cal K}_{\mu\nu} - {\cal K}\gamma_{\mu\nu} -\frac{2}{\ell}\gamma_{\mu\nu} \right] -\frac{1}{\ell}|\psi|^2\gamma_{\mu\nu} \nn
J_\mu		&=& \frac{1}{\sqrt{-\gamma}}\frac{\delta S_{\rm ren}}{\delta A^\mu} =n^M F_{M\mu}\,,
\eea
are finite in the limit $r\rightarrow \infty$. An explicit calculation reveals that the background solution reproduces the stress tensor and current of an ideal conformal\footnote{One can show that ${\rm Tr} T_{\mu\nu}\propto \xi_1\xi_2$, so that the fluid satisfies the tracelessness requirement of conformal invariance if and only if either $\xi_1$ or $\xi_2$ is zero. In the dual field theory this corresponds to {\it not} deforming by the relevant operator ${\cal O}_\psi$. Of course finite temperature and/or finite chemical potential then break conformal invariance.} relativistic fluid, which we can see from the leading order behaviour
\be\label{eq:zero_order_stress}
\lim_{r\rightarrow \infty} \frac{r}{\ell} T_{\mu\nu} = \frac{\varepsilon}{2}\left( \eta_{\mu\nu}  + 3u_\mu u_\nu \right) \,,\qquad \lim_{r\rightarrow\infty} \frac{r}{\ell} J_\mu = \rho u_\mu 
\ee
We can also determine the total energy of the solution,
\be
E = \int d^2 x \frac{r}{\ell} T_{00} =  \varepsilon\,  {\rm vol}_2\,,
\ee
where  the total (divergent) two-volume ${\rm vol}_2$ comes from the integral over spatial directions.
\subsection{Thermodynamics}
Having already determined the counterterms, we can use standard Euclidean techniques \cite{Gibbons:1976ue} to find the thermodynamics of this system. We want to calculate the Euclidean action $I_E$, which takes on the r\^ole of the Gibbs free energy for the holographic superfluid in the saddle point approximation. The results in this section are standard so we will give few details on their derivation. We take
\be
v = -i\tau\,, \qquad I_E = -i S_{\rm ren}\,.
\ee
and use the fact that the gravitational Euclidean partition function defines a thermodynamic potential via
\be
\ln {\cal Z} = -\beta \Omega\,.
\ee
Following the same kind of manipulations as in \cite{Gauntlett:2009bh,Gauntlett:2009dn} we find that the manifestly finite Euclidean action gives us the thermodynamic potential
\be
I_E = \beta \Omega(\mu,T) = \frac{{\rm vol}_2}{T} \left(  \varepsilon - \mu \rho- T s \right)\,,
\ee
where $s$ is entropy density (the entropy $S = \frac{A}{4G}$ is defined as usual in terms of the horizon area).
From the on-shell variation of the action we learn that the action is stationary for fixed temperature $\beta^{-1}$ and fixed chemical potential $\mu$. Furthermore, stationarity of the action requires that either $\xi_1=0$ or $\xi_2=0$. Thus we have the first law
\be\label{eq:first_law}
\delta \varepsilon = T \delta s + \mu \delta \rho \,.
\ee
We will now generalise this system by  adding an additional chemical potential, related to the Goldstone field of the broken global $U(1)$ of the boundary theory. This has the physical interpretation of allowing for a finite supercurrent to flow in the broken phase. The bulk manifestation of this is a non-zero value for the spatial components of the vector field, best quantified by saying that the gauge invariant expression $A_\mu - \partial_\mu \alpha $ develops a non-trivial value in the bulk.

\section{Introducing a Supercurrent}
Let $n_\mu$ be a constant vector of unit magnitude, such that 
\be\label{eq:bulk_joseph}
\gamma_{\infty}^{\mu\nu}n_\mu u_\nu=0
\ee
This means that in a frame where $u^0= 1\,, \mathbf{u}=0$, we can write $n^0=0$ and the spatial components are given by $\mathbf{n}$,  a unit two vector.
 We consider the bulk ansatz
\bea\label{eq:BulkMetric}
ds^2 &=& ds^2_{\rm static}  + \frac{r^2}{\ell^2}\, \left( 2  \, {\cal C} u_{(\mu}n_{\nu)}  - {\cal B}  n_\mu n_\nu \right) dx^\mu dx^\nu + \frac{2{\cal C}h}{f}n_{\mu} dx^\mu dr \nn
A &=& \left(-\phi u_\mu + \Phi n_\mu\right) dx^\mu - \frac{\phi h}{r^2 f}dr \nn
\psi &=& \xi\,  e^{iq\alpha} \,.
\eea
The additional functions in this ansatz depend only on the radial direction. The condition (\ref{eq:bulk_joseph}) can be imposed without loss of generality in the sense that any constant {\it space-like} vector $n_\mu$ defines a bulk metric that is diffeomorphically equivalent to (\ref{eq:BulkMetric}).  So, by splitting up such a space-like vector into components parallel to $u^\mu$ and components normal to $u^\mu$ we can arrange for (\ref{eq:bulk_joseph}) to hold. 

For specific calculations we will always work in a gauge where the scalar $\psi$ is real. However, notice that such solutions are gauge-equivalent to ones where the phase $\alpha$ has a boundary dependence that is linear in boundary coordinates. To see this, one applies a gauge transformation
\be
A_\mu \rightarrow A_\mu + \partial_\mu (  x^\nu\lambda_\nu)  \,,\qquad \alpha \rightarrow \alpha +  x^\nu\lambda_\nu\,.
\ee
Let us define the combination
\be
\alpha_M = A_M - \nabla_M\alpha \,.
\ee
Then the quantity
\be
\alpha_\mu^\infty = \lim_{r\rightarrow\infty} \left(A_\mu - \partial_\mu \alpha \right)
\ee
is gauge invariant with respect to the bulk U(1) and therefore an operationally convenient measure of the magnitude of the boundary Goldstone mode. We note that
\be\label{eq:Goldstone}
\alpha_\mu^\infty \alpha^{\mu\,\infty} = -\mu^2 + \Phi_\infty^2
\ee
gives us a gauge-invariant definition of the boundary supercurrent density.
The latter part of this paper will be concerned with numerically solving the Einstein-Scalar-Maxwell equations \bea
G^M_N - \frac{3}{\ell^2} \delta^M_N &=& \frac{1}{2}F_{NP}F^{MP} - \frac{1}{2}\delta^M_N\left(\frac{1}{4}F^2   +  V  + \left(  (\nabla\xi)^2  + q^2 \xi^2 \alpha^2 \right)\right)\nn
& & + \left( q^2 \alpha^M\alpha_N \xi^2 + \nabla^M \xi \nabla_N \xi  \right)\nn
\Box \xi &=&q^2 \alpha^2 \xi + \frac{1}{2}V'\nn
\nabla_M F^M{}_N &=& -2 q^2 \xi^2 \alpha_N\nn
\nabla_M\left( \xi^2 \alpha^M \right) &=& 0
\eea
with the above ansatz. They are given explicitly in appendix A.
 However, even without numerics,  we can learn a lot about the system by analysing its behaviour in an analytic near-boundary expansion. Let us turn to this analysis now.

\subsection{Asymptotic analysis}

The equations allow the following asymptotic expansion (the remaining fields have the same form as in (\ref{eq:expansion_one}) ):
\bea\label{eq:AsymptoticFields}
 f &=& h_\infty^2 - \frac{8 \pi G \ell^2 \varepsilon}{ r^3}  -\frac{8 \pi G \ell^2 {\cal B}^{(3)}  }{3r^3} + \cdots  \nn
{\cal B} &=& {\cal B}^{(0)} + \frac{16 \pi G \ell^4 {\cal B}^{(3)}}{3 r^3} + \cdots \nn
{\cal C} &=& {\cal C}^{(0)} + \frac{16 \pi G \ell^4{\cal C}^{(3)}}{3 r^3} + \cdots  \nn
\Phi  &=& \left(\mu_s - \frac{J_s \ell^2}{r} + \cdots \right)\,.
\eea
In order to keep the asymptotic boundary metric in Minkowski form, we demand that the non-normalisable parts of ${\cal B}$ and ${\cal C}$ vanish. Thus we restrict attention to solutions with
\be\label{eq:NonNorm}
 {\cal B}^{(0)} =  {\cal C}^{(0)} = 0\,.
\ee

\subsection{Free energy and First Law}\label{sec:thermo}
We can now identify the relevant thermodynamic potentials and variables of the boundary theory from our asymptotic bulk analysis.
In order to relate the Euclidean action to a thermodynamic partition function we first have to write the bulk part of the action as a total derivative. This can be done in two different ways:
\bea
S_{\rm bulk}^{(I)} &=& \frac{i \beta{\rm vol}_2}{16 \pi G} \int_{r_+}^\infty \left( \frac{2}{r}\sqrt{-g}g^{rr} \right)' dr\nn 
S_{\rm bulk}^{(II)} &=& i \beta{\rm vol}_2\int_{r_+}^\infty \left[\frac{1}{16\pi G}\frac{r^4}{\sqrt{-g}}\left( g^{rr} \frac{g}{r^4} \right)'   + \sqrt{-g} g^{rr} F_{r\mu}A^\mu\right]'dr\,.
\eea
The first expression gets contributions only from the upper boundary of the integral, whereas the second expression receives both horizon and boundary contributions.

Let us consider the on-shell variation of the Euclidean action. This receives no contribution from the lower boundary of the integral. From the upper boundary we find for variations that preserve the (conformally) Minkowski metric on the boundary:

\be\label{eq:stationary_action}
\delta I^{\rm bulk}_E + \delta I^{\rm counter}_E \Bigr|_{\rm OS} =\beta {\rm vol}_2 \left[\left(\varepsilon -\mu\rho \right)\delta h_\infty   - \rho\delta\mu + J_s\delta\mu_s \right]\,.
\ee
For fixed $\beta$, we have $\delta h_\infty = - \frac{\delta T}{h_\infty T}$, so that the action is stationary for fixed $\mu,\mu_s,T$. Putting all this together we learn that the Euclidean action defines a grand canonical ensemble in terms of temperature $T$ and chemical potentials $\mu$ and $\mu_s$.

 Using
\be\label{eq:ensemble}
I_{E}^{\rm bulk} + I^{\rm counter}_E = \beta \Omega(\mu,\mu_s,T) = \beta {\rm vol}_2\, \omega(\mu, \mu_s,T)
\ee
we find the quantum statistical relation
\bea
\omega^{(I)}(\mu, \mu_s,T) &=& -\frac{\varepsilon}{2} -  \frac{{\cal B}^{(3)}}{2} = -P\,,
\eea
where $P$ is pressure, following from the first form of the action, and
\bea\label{eq:quantum_statistical}					
\omega^{(II)}(\mu, \mu_s,T) &=& 2 P- Ts- \mu\rho +  \mu_s J_s 
\eea
from the second form. Equating the two different expressions for the thermodynamic free energy gives rise to a Smarr-Gibbs-Duhem relation:
\be\label{eq:smarrgibbsduhem}
3 P = Ts+ \mu\rho - \mu_s J_s \,.
\ee

Using (\ref{eq:quantum_statistical}) and (\ref{eq:stationary_action}) we can deduce the first law in the form
\be
 \delta \omega = -\left(s + \frac{\varepsilon - 2P}{T} - \frac{\mu_s J_s}{T} \right)\delta T  - \rho \delta \mu + J_s \delta \mu_s 
 \ee
Note that via (\ref{eq:Goldstone}) and (\ref{eq:AsymptoticFields}) the third term involves a variation of the norm of the boundary Goldstone field.
Using the relation $\omega = -P$ we can rewrite this in the form
\bea
\delta P 	&=& \hat s \delta T + \rho_n \delta\mu - \frac{\rho_s}{2\mu}  \delta \left( (\alpha_\mu^\infty)^2  \right)
\eea
where we defined the normal and superfluid density as
\be\label{eq:SuperfluidDensityI}
\rho = \rho_n + \rho_s\,, \qquad{\rm with}\qquad \rho_s = \frac{\mu J_s}{\mu_s}
\ee
The conjugate function of temperature, 
\be\label{eq:entropywithcontribs}
\hat s = s + \frac{\varepsilon - 2P - \mu_s J_s}{T}\,,
\ee
 is the thermodynamic entropy, which gets contributions not only from the horizon entropy, but also from the hair of the black hole. We will see in the next section that these contributions in fact cancel each other, so that the final expression of the entropy density is that of the black-hole horizon. Thus the pressure $P\left( T,\mu,(\alpha_\mu)^2 \right)$  is the same thermodynamic potential as that appearing in the context of relativistic field theory in \cite{Son:2000ht,Son:2000ht2}. Here we gave a derivation from bulk gravity, which via the AdS/CFT correspondence is dual to a strongly-coupled relativistic field theory on the boundary.
We can rewrite the Smarr-Gibbs-Duhem relation as
\bea
P + \epsilon &=& T\hat s + \mu \rho_n\,.
\eea
It is obvious from this definition that $\epsilon$ is the Legendre transform of pressure with respect to temperature and chemical potential. From (\ref{eq:smarrgibbsduhem}) and (\ref{eq:SuperfluidDensityI}), it is related to the zero-zero component of the energy-momentum tensor via
\be
\epsilon = \varepsilon - \mu \rho_s\,.
\ee
\subsection{The energy momentum tensor and current}\label{sec:emtensor}
Using the renormalised expressions (\ref{eq:StressAndCurrent}) we can deduce the form of the boundary stress tensor and current following from the bulk ansatz (\ref{eq:BulkMetric}). We find
\bea\label{eq:superfluid_constitutive}
T_{\mu\nu} &=&(\varepsilon + P)u_\mu u_\nu + P \eta_{\mu\nu} - {\cal B}^{(3)}n_\mu n_\nu+ 2 {\cal C}^{(3)}u_{(\mu}n_{\nu)}\nn
J_\mu &=&\rho u_\mu - J_s n_\mu\,.
\eea
Notice that the stress tensor is traceless on account of the orthogonality relation between $n^\mu$ and $u^\mu$ and the definition of pressure in terms of $\varepsilon$ and ${\cal B}^{(3)}$. One might be a little surprised by the occurrence of a cross term between $u^\mu$ and $n^\mu$. To elucidate this point, let us compare our form of the constitutive relations with those of \cite{Herzog:2009md,Son:2000ht}. These in turn were shown to be equivalent to the original formulation of Israel, Carter, Khalatnikov, and Lebedev (see references in  \cite{Herzog:2009md,Son:2000ht}).

Let us define a new superfluid velocity $v^\mu$ via
\be
n_\mu = \frac{{\cal C}^{(3)}}{{\cal B}^{(3)}}\left( u_\mu - v_\mu \right)\qquad {\rm with} \qquad u^\mu v_\mu = -1
\ee
In the new frame we get a stress tensor and current
\bea\label{eq:newFrame}
T_{\mu\nu} &=&\left( \varepsilon +  \frac{[{\cal C}^{(3)}]^2}{{\cal B}^{(3)}}+ P\right)u_\mu u_\nu + P \eta_{\mu\nu} - \frac{[{\cal C}^{(3)}]^2}{{\cal B}^{(3)}}v_\mu  v_\nu \nn
J_\mu &=&\left( \rho - J_s\frac{{\cal C}^{(3)}}{{\cal B}^{(3)}}  \right)u_\mu + J_s\frac{{\cal C}^{(3)}}{{\cal B}^{(3)}}v_\mu
\eea
The coefficient of $v^\mu$ is the superfluid density $\rho_s$. From section \ref{sec:thermo} the entropy density in the fluid rest frame is $\hat s$. To get to the arbitrary reference frame of (\ref{eq:newFrame}) we must apply the boost $u^\mu$. From this, it follows that the current $\hat s u^\mu$ describes the flow of entropy.  As in \cite{Landau:1941}, one can constrain the form of the constitutive relations by requiring a vanishing divergence of the entropy current
\be
\hat s^\mu = \hat s u^\mu
\ee
This requires that
\be\label{eq:cond1}
\mu J_s = - {\cal C}^{(3)}\qquad {\rm with}\qquad \rho_s = J_s\frac{{\cal C}^{(3)}}{{\cal B}^{(3)}}
\ee
Finally, we should compare this expression for the superfluid density to that in Eq. (\ref{eq:SuperfluidDensityI}). Evidently we must have
\be\label{eq:cond2}
\frac{\mu}{\mu_s} = \frac{{\cal C}^{(3)}}{{\cal B}^{(3)}}\,.
\ee 
Interestingly, using (\ref{eq:cond2}) and (\ref{eq:cond1}) in (\ref{eq:entropywithcontribs}) we see that the additional contributions to the entropy cancel against each other, so that $\hat s = s$.
We just derived these constraints among the boundary data from a careful analysis of the thermodynamics following from a semiclassical treatment of the bulk quantum partition function. The conditions (\ref{eq:cond1}) and (\ref{eq:cond2}) have to be satisfied if the boundary is to have a sensible thermodynamic description and indeed they imply that the entropy $\hat s$ is in fact the thermodynamic entropy $s$ of the black hole. However, we cannot simply impose these constraints on the asymptotic data of any generic solution. There is not enough freedom in the equations to do this. We therefore need to check that the full bulk solutions, once appropriate regularity conditions have been given at the horizon, satisfy these constraints {\it automatically}.
We will see below that both relations between asymptotic data coefficients are satisfied by our numerical solutions to very high accuracy (see figure \ref{fig:Constraints}). We should see this result as a non-trivial consequence of the interplay between horizon regularity and boundary fluid dynamics that can only be seen at the full non-linear level.

The description of superfluids we have derived in the preceding sections is identical to that proposed by Landau \cite{Landau:1941}, and later generalised to the relativistic domain by Khalatnikov et al.,  \cite{LLvol6}.

Note that when going on to include dissipative terms, {\it i.e.} terms of higher order in the derivative expansion, the conservation of the entropy current should be replaced by the requirement that its divergence be non-negative. A positive entropy current has been constructed for general fluid-dynamical solutions without bulk symmetry breaking in \cite{Bhattacharyya:2008xc} and it would be interesting to see if such a construction could also be carried out in our case. The analysis in this paper establishes semi-analytically - in the sense that the horizon-to-boundary map had to be constructed numerically - that such a current exists at non-dissipative order.

Finally note that in terms of $v^\mu$, the gauge invariant quantity evaluates to
\be
\alpha_\mu^\infty= -\mu v_\mu\,.
\ee
By choosing a different gauge in the bulk, we can make direct contact with the standard treatment, given for example in \cite{Son:2000ht}. Thus, if we take $\partial_\mu\alpha = \mu  v_\mu ,$
the Josephson condition becomes
\be
\mu  + u^\mu\partial_\mu\alpha =0\,,
\ee
where now $\mu$ is the chemical potential appearing in  \cite{Son:2000ht}. Note that we had to use the constraints (\ref{eq:cond1}) and (\ref{eq:cond2}), which we have confirmed numerically,  in order to connect the quantities $n_\mu$ and consequently $v_\mu$ to the superfluid velocity. Analytically these relations follow from requiring a positive semi-definite divergence of the entropy current. However this property of the divergence should follow from the second law of black-hole thermodynamics and establishing this law in the present context would constitute an analytical proof of the identifications of (\ref{eq:cond1}) and (\ref{eq:cond2}).

\subsection{Local bulk symmetries $\rightarrow$ Boundary hydrodynamics}
One use for the solutions we have constructed is as a homogeneous starting point for a hydrodynamic gradient expansion of the form pioneered in \cite{Bhattacharyya:2008jc}. Although many interesting physical properties have been obtained in (ordinary) perturbative studies \cite{ Herzog:2008he,Basu:2008st,Herzog:2009md}, these perturbative solutions do not serve as a good starting point for a gradient expansion and we comment on this further in the discussion section. We hope to return to the dissipative order of the gradient expansion in the future. However, even without explicitly carrying out the higher orders of this expansion, we can already deduce the hydrodynamic relations these will satisfy from general symmetry principles.

Among the Einstein equations there are constraint equations, which are first order in radial derivatives, and dynamical equations, which are of second order. In the context of a gradient expansion of the universal gravity subsector of AdS/CFT it was shown in \cite{Bhattacharyya:2008jc} that these constraint equations at order $n$ imply that the boundary stress tensor and $U(1)$ current satisfy the equations of fluid dynamics up to order $n-1$. However, these equations in fact follow from elementary symmetry considerations. Namely, boundary current conservation follows from the $U(1)$ gauge symmetry in the bulk, and the remaining equations, involving the stress-energy tensor, can be derived by considering an infinitesimal diffeomorphism generated by a vector field lying parallel to the boundary directions.

The resulting equations take the form of a conservation equation for the current and an equation of continuity for the stress-energy tensor:
\be\label{eq:fluid_cons}
\partial_\mu J^\mu = 0\,,\qquad \partial_\mu T^{\mu\nu} = {\cal F}^{\nu}{}_{\lambda}J^\lambda
\ee
where ${\cal F}$ is the applied {\it boundary} field strength tensor (to be distinguished from the bulk field strength $F$). The bulk gradient expansion then serves to derive the form of the conserved stress and current order by order. In this work we have determined all non-dissipative contributions. It is a straightforward extension to go on to include dissipative terms.

 From the bulk analysis we have also learned that
\be\label{eq:Josephson}
v^\mu u_\mu =-1\,,
\ee
which gives us the `Josephson equation'  governing the dynamics of the Goldstone mode. From a field-theory perspective it has been elaborated upon in \cite{Son:2000ht}. Together with the constitutive relations (\ref{eq:superfluid_constitutive}), the equations (\ref{eq:fluid_cons}) and (\ref{eq:Josephson}) form the full set of equations of conformal relativistic superfluid hydrodynamics.
This is as far as the analytical results can take us. We will now turn to the numerical part of this paper.

\section{Numerical Results}

 \begin{figure}[t!]
\begin{center}
\vskip3em
\includegraphics[width=0.95\textwidth]{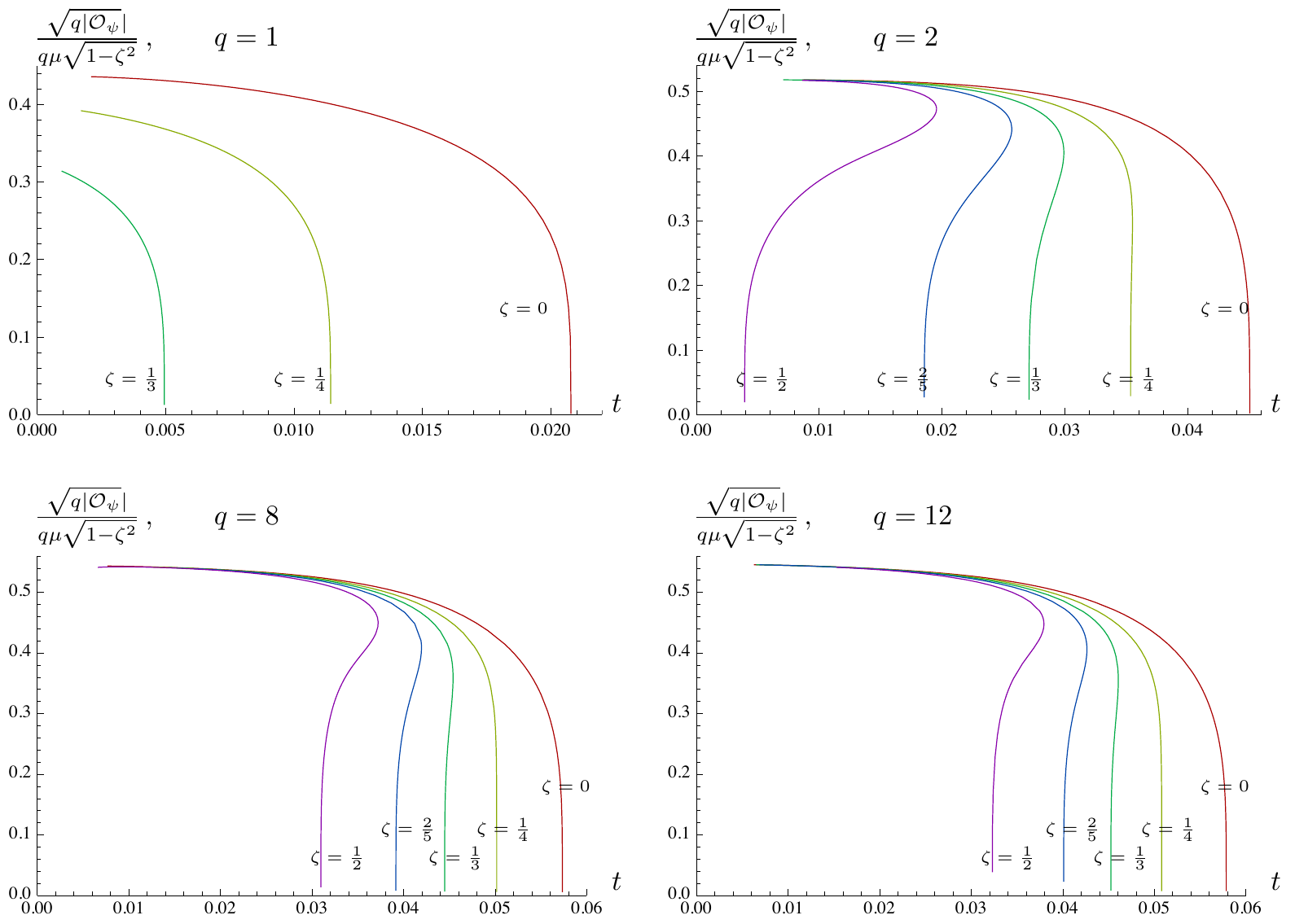}
\caption{\it (colour online) The condensate $\langle {\cal O}_\psi\rangle$ as a function of reduced temperature. For larger values of the charge $q$ the condensate becomes multi-valued beyond a critical fraction of superfluid $\zeta_c$ indicating that the phase transition is now first order. The normalisation of the condensate is chosen for ease of comparison with the probe calculations of \cite{Herzog:2008he}. 
\label{fig:Condensate}}
\end{center}\end{figure}
 
We wish to solve the differential equations of appendix \ref{app:bulkequations} subject to the following two requirements
\begin{enumerate}
\item There is a {\it regular} horizon in the bulk at $r=r_+$. 
\item The asymptotic boundary metric remains conformally flat.
\end{enumerate}
This is the minimal set of physical requirements that ensures that the boundary theory is dual to a locally equilibrated fluid living on Minkowski space. From the second requirement we already deduced the condition (\ref{eq:NonNorm}) on the asymptotic expansion (\ref{eq:AsymptoticFields}). It is convenient to choose units (that is using a scaling symmetry) to fix $r_+ =1$.  While the second condition has already been incorporated in the asymptotic boundary expansion, the first requirement fixes the near-horizon expansion to take the form
\bea
f(r) &=& f_+ (r-r_+) + \cdots\nn
h(r) &=& h_+ + \cdots \nn
{\cal B}(r) &=& {\cal B}_+ + \cdots \nn
{\cal C}(r) &=& {\cal C}_+ (r-r_+) + \cdots \nn
\phi(r) &=& \phi_+ (r-r_+)\nn
\Phi(r) &=& \Phi_+ + \cdots\nn
\xi(r) &=& \xi_+ + \cdots\nn
&{\rm with}&\nn
f_+ &=& h_+^2 \left( 3 + 16 \pi G \xi_+^2 \right) - 4 \pi G \ell^2 \phi_+\,,
\eea
where $\{ h_+, {\cal B}_+, {\cal C}_+,\phi_+,\Phi_+,\xi_+  \}$ are the independent horizon data and $f_+$ and all higher-order coefficients are uniquely determined by them. Regularity demands that ${\cal C}$ vanishes at the horizon in such a way that $\frac{\cal C}{f}$ remains finite, as can be seen for instance by inspecting the $n_\mu dx^\mu dr$ term of the metric in (\ref{eq:BulkMetric}). Similarly $\phi(r_+) = 0$ for regularity.
In practise we determine the expansion to a high order and use it to set boundary conditions near the horizon.

The total differential order of the system of ODEs we are solving is twelve. After taking account of all scaling symmetries we are left with eight more data at infinity, $\{ {\cal B}^{(3)}, {\cal C}^{(3)},\mu,\rho,\mu_s, J_s,h_\infty, \varepsilon \}$, giving a total of  $14$ independent pieces of data. We will thus get a two-parameter family of solutions. Of course we are free to choose any two non-conflicting parameters, but a natural and convenient choice is the temperature $T$ and the superfluid chemical potential $\mu_s$. Because of the underlying conformal symmetry, all physical quantities will depend on the dimensionless ratios $\frac{T}{\mu}$ and $\zeta = \frac{\mu_s}{\mu}$. We will refer to $t = \frac{T}{q\mu \sqrt{1-\zeta^2}}$ as the {\it reduced temperature} and $\zeta$ as the {\it superfluid fraction}.

\subsection{The condensate}

\begin{figure}[t!]
 \vskip3em
\begin{center}
\includegraphics[width=0.95\textwidth]{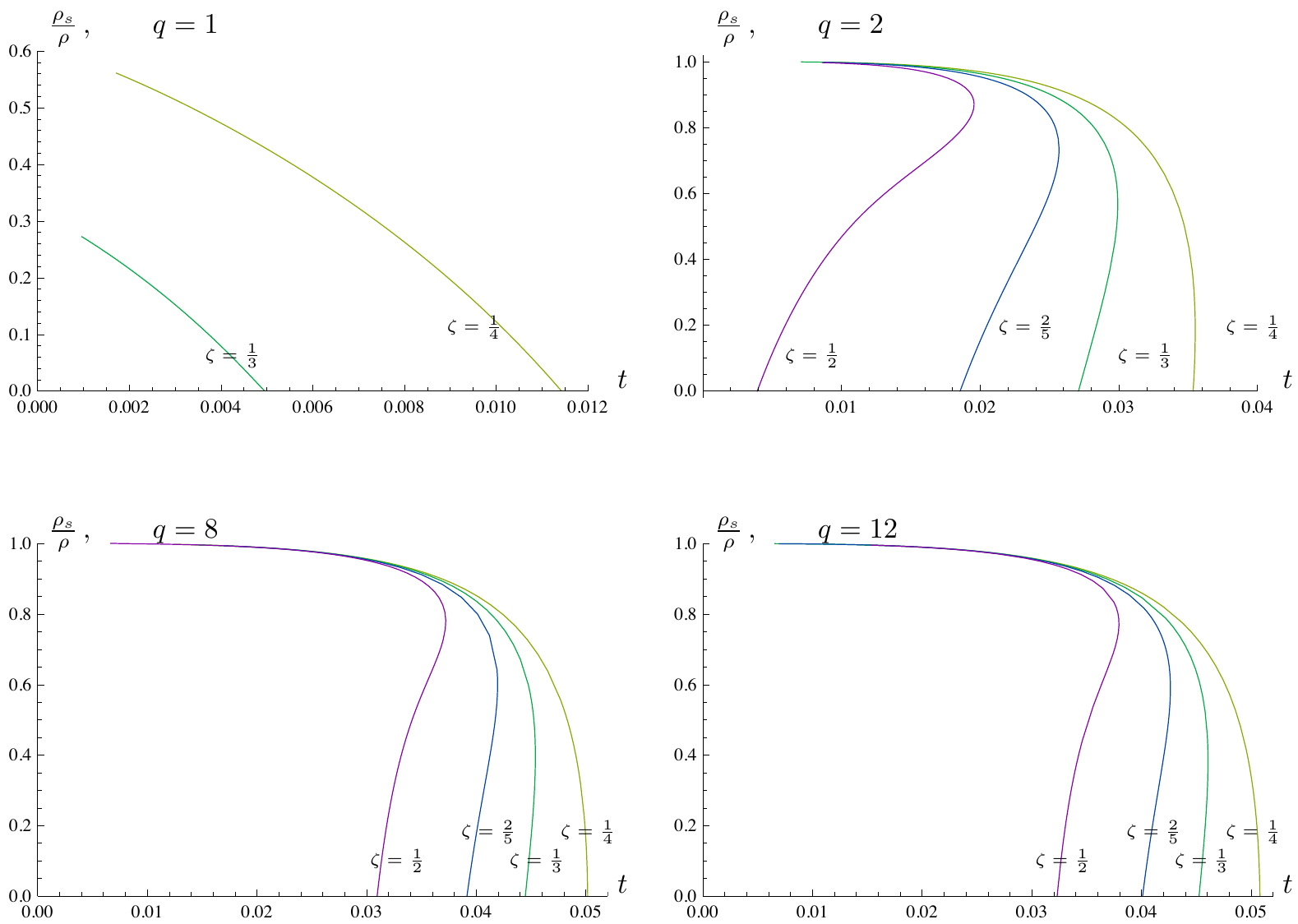}
\caption{\it (colour online) The quantity $\rho_s/\rho$ as a function of reduced temperature. This fraction approaches unity for $q > 1$, but for $q=1$ there always remains a normal component, even at very low temperatures. Some of the curves are multi-valued as a consequence of the first-order phase transition. In those cases only the upper branch below the critical temperature $t_c(\zeta)$ is physical.
\label{fig:SuperfluidComponent}}

\end{center}
\vskip1em
\end{figure}

The $r^{-2}$ asymptotic behaviour of the field $\psi$ sets the expectation value of the symmetry-breaking dimension-two operator ${\cal O}_\psi$ in the boundary theory.
\be
 \langle {\cal O}_\psi \rangle = \sqrt{2}\, \xi_2
\ee
This operator tends to condense at low temperatures. By numerically constructing $\xi_2$ we determined the graphs shown in figure \ref{fig:Condensate} illustrating the dependence of the condensate on reduced temperature and superfluid fraction.  There is always a critical superfluid fraction above which the system is forced into its normal state. The same behaviour was found in probe-limit calculations \cite{Herzog:2008he,Basu:2008st}, as well as for rotating superfluids in \cite{Sonner:2009fk}. Furthermore, for sufficiently high charge $q$ of the field dual to the condensing operator there is also a special point on the phase diagram below which the superfluid transition is first order and above which it is second order. This behaviour was already found in the probe limit calculations. However, we find that for lower values of $q$ the transition never becomes first order. This is clearly a result of strong gravitational backreaction. In order to conclusively show that the transition is first order or second order we also computed the free energy as a function of temperature for fixed $\zeta$.

\subsection{Free energy and superfluid density} 

\begin{figure}[b!]
 \vskip3em
\begin{center}
\includegraphics[width=0.95\textwidth]{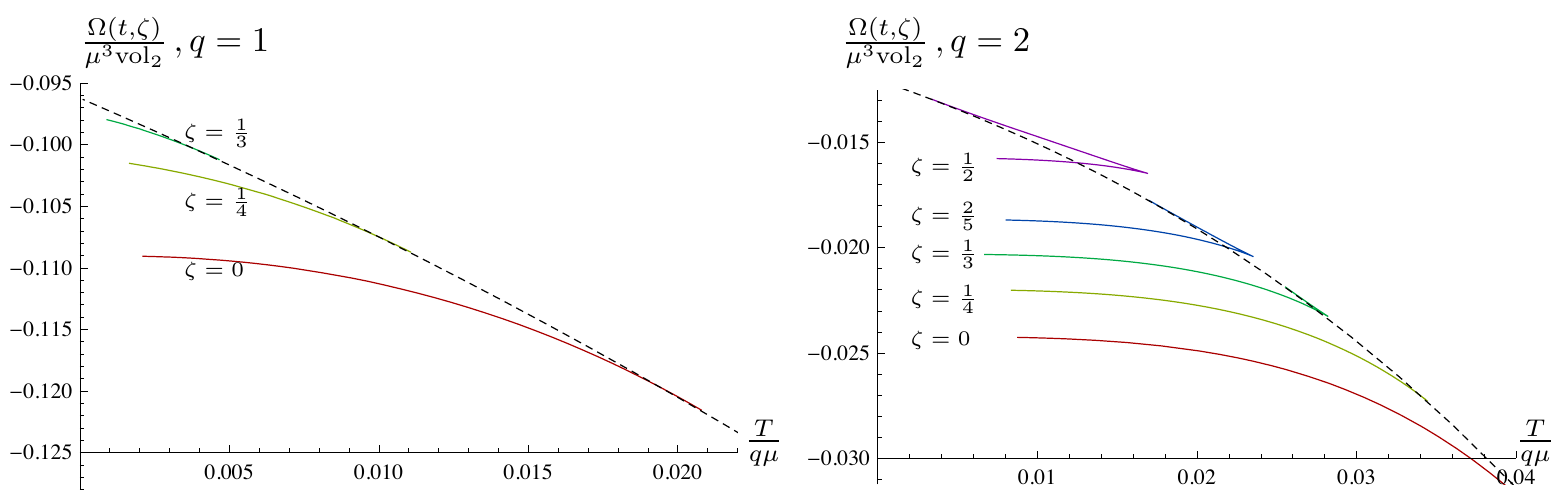}
\caption{\it (colour online) The free energy at fixed $\zeta$ as a function of reduced temperature for $q=1,2$. The  unbroken branch is shown as a dashed line. For the lowest $q$ the broken branch joins the unbroken branch smoothly for all allowed values of $\zeta$. For $q=2$, we see the swallow-tail behaviour characteristic of first-order transitions for fractions above a critical value $\zeta_c$. The behaviour for $q>2$ is very similar and we do not reproduce these plots here.
\label{fig:FreeEnergy}}
\end{center}
\vskip1em
\end{figure}

Figure \ref{fig:FreeEnergy} show numerical graphs of the thermodynamic free energy
\be
\frac{\Omega\left( T,\zeta_0 \right)}{\mu^3 {\rm vol}_2} \qquad {\rm with}\qquad S = -\frac{\partial \Omega\left( T,\zeta_0 \right)}{\partial T} \Biggr|_{\mu,\mu_s}
\ee
 at fixed superfluid fraction.
We show curves for $q=1,2$. For sufficiently high $q$  and larger fractions $\zeta$, we clearly see that the smooth second-order behaviour goes over into the swallow-tail cusp behaviour characteristic of first-order phase transitions, but for $q=1$ the free energy remains smooth and the transition second order up until the critical value $\zeta_c$ above which no superfluid phase exists. The free energy higher values of $q$ looks very similar to the $q=2$ case and we do not show the corresponding plots here. The bulk charge $q$ appears in the boundary theory as a normalisation factor of a ratio of three-point and two-point functions \cite{Gubser:2008px}, so from this perspective there appears to be no need to restrict its value to be above a certain treshold or indeed to be integer valued. We use integer values in our numerics only for convenience, as is the case in most of the pertinent literature. Figure \ref{fig:SuperfluidComponent} presents plots of the behaviour of $\rho_n$ and $\rho_s$ as a function of temperature. The double-valued nature of the graphs is a consequence of the first-order phase transition. At $T_c$ the system spontaneously develops a {\it finite} value of the condensate and immediately jumps to the thermodynamically preferred branch. Again we emphasise that the low-temperature behaviour involves strong backreaction and can therefore not be determined in the probe limit.

\begin{figure}[t!]
\begin{center}
\includegraphics[width=0.95\textwidth]{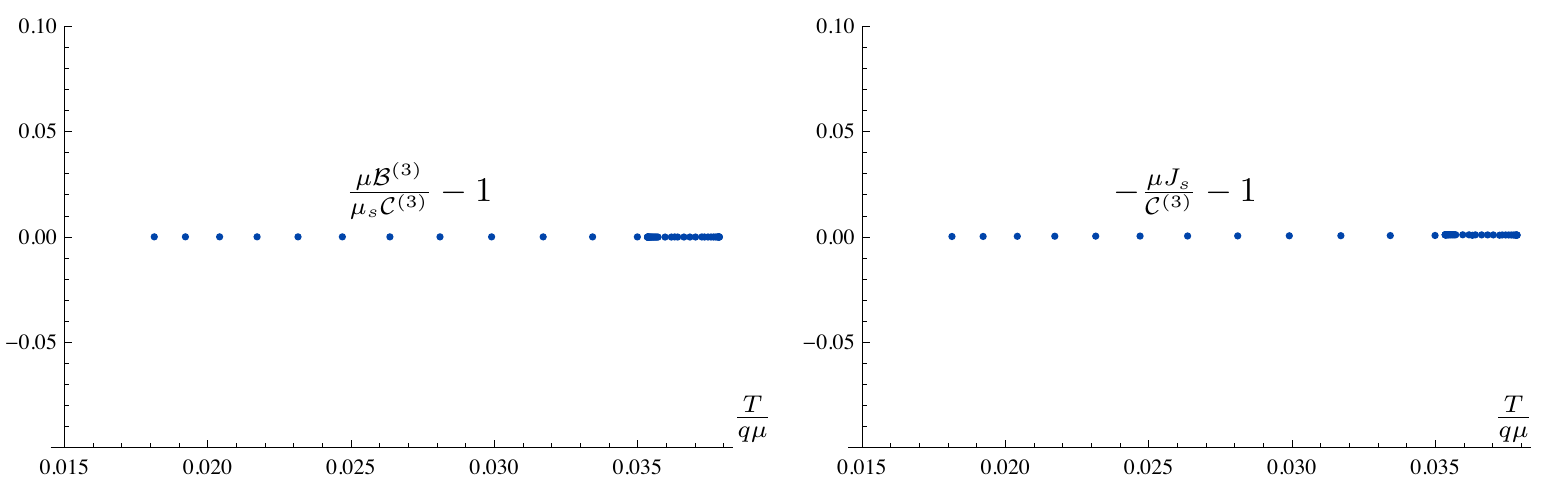}
\caption{\it (colour online)  Figure demonstrating to numerical accuracy (in this case the numerical error is of the order of $10^{-4}$) the constraints among asymptotic data following from horizon regularity. Both constraints are illustrated based on the data points we used to generate the plots at $q=8$ and $\zeta=\frac{2}{5}$. \label{fig:Constraints}}
\end{center}\end{figure}

Figure \ref{fig:Constraints} demonstrates the validity, to numerical accuracy, of the relations between the two expressions for the superfluid density, Eqs. (\ref{eq:cond1}) and (\ref{eq:cond2}), that we derived in section \ref{sec:emtensor} from basic thermodynamic considerations.

\begin{figure}
\begin{center}
\includegraphics[width=0.6\textwidth]{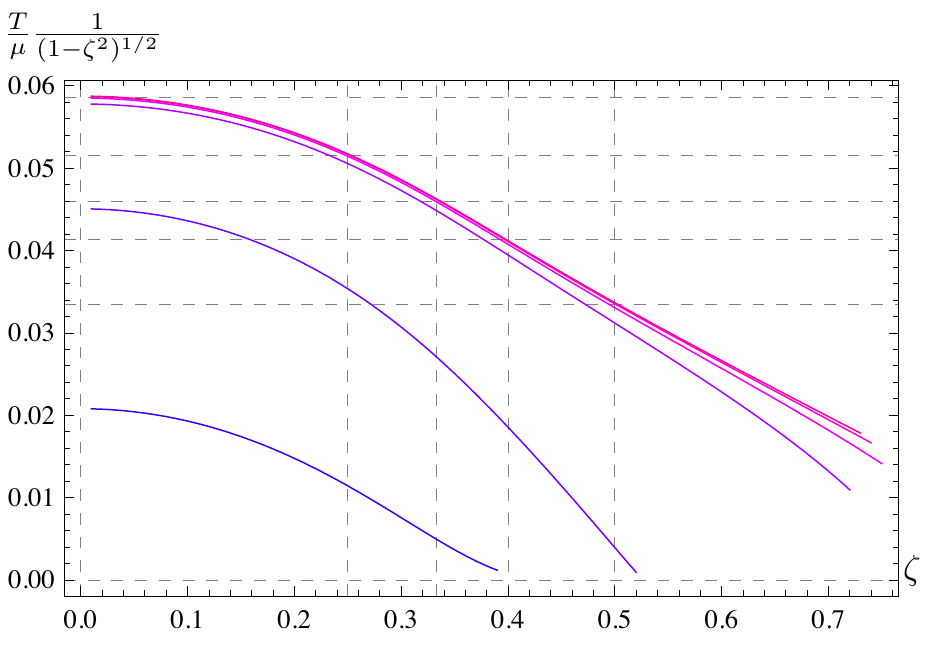}
\caption{\it (colour online)  Temperature, at which the marginal mode occurs as a function of superfluid fraction $\zeta$. The curves for ascending $q=1,2,8,16,32,100$ are shown from bottom to top. The vertical dashed lines indicate fractions of the condensate of $\zeta=1/4, 1/3, 2/5, 1/2$ and the horizontal dashed lines are visual aids in order to identify the critical temperatures for the $q=100$ curve.\label{fig:Tcdiagram}}
\end{center}\end{figure}

Finally, in figure \ref{fig:Tcdiagram} we show that the critical temperatures of \cite{Herzog:2008he} are recovered for large $q$ for appropriately rescaled quantities. Note that even though $q=1,2$ are far from the probe limit, $q=8$ already gives a close approximation to the critical temperature in the probe limit.

We shall conclude with a brief discussion of the relevance of our approach to determining higher-order transport properties of strongly coupled relativistic superfluids and  mention some open questions. 
\section{Discussion}
If one attempts to formulate a gradient expansion of a hydrodynamical system all transport coefficients at non-zero derivative order are subject to an ambiguity stemming from the freedom to shift zeroth-order quantities by gradients. In the gravity setup these shifts manifest themselves as homogeneous modes of the bulk equations. One can resolve this ambiguity by fixing appropriate frames, a popular choice being the Landau frame.
 Alternatively one can concentrate on a set of scheme-independent quantities. However, both approaches show that the set of scheme-independent quantities (transport coefficients) at first order, and therefore in turn all higher orders, depends on the number of independent thermodynamic quantities present already at zeroth order. This means that, for a well-defined gradient expansion that includes non-trivial transport coefficients associated to the superfluid component, we must incorporate the superfluid velocity in the homogeneous zeroth-order solution. We have constructed such a supercurrent solution in this paper. Incorporating the supercurrent only perturbatively at zeroth order \cite{Herzog:2009md} means that the hydrodynamical expansion must be a double expansion, both in slow gradients and small superfluid density. This is appropriate for determining physical quantities, such as various sound speeds away from low temperature or large condensates \cite{Karch:2008fa,Yarom:2009uq}, but we would expect the method to break down at some point, when the superfluid component strongly backreacts on the metric. It is thus an interesting project to determine systematically, using our techniques, the new transport coefficients associated with the superfluid component to first and perhaps second order.
 
\begin{figure}[t!]
\begin{center}
\includegraphics[width=1.0\textwidth]{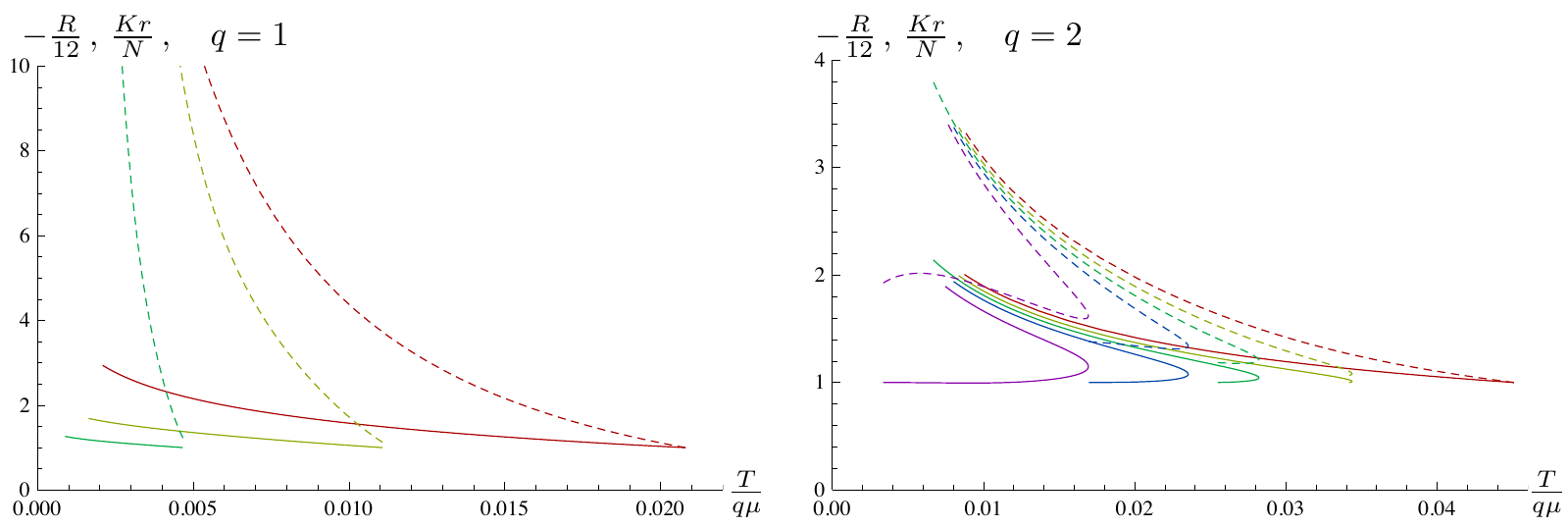}
\caption{\it (colour online)  Curvature at the horizon for different fractions of superfluid. The solid lines depict the Ricci scalar as a function of temperature and the dashed lines the Kretschmann scalar $Kr=\sqrt{R_{MNPQ}R^{MNPQ}}$, whose normalisation $N$ is chosen so that $Kr=N$ at $t_c(\zeta=0)$. Both quantities diverge at very low temperatures, indicating that the supergravity approximation breaks down.
\label{fig:Curvature}}
\end{center}\end{figure}

This work assumes that classical (super) gravity is a good approximation to the underlying string theory. This is true so long as the geometric curvature scales are small and thus the dual gauge theory is strongly coupled. Figure \ref{fig:Curvature}  demonstrates that the assumption that curvatures remain small is not valid at low temperatures close to the horizon for $q=1,2$ (again, the qualitative behaviour for $q>2$ is similar to $q=2$ and thus not shown).  This was demonstrated previously for zero superfluid fraction in \cite{Gauntlett:2009dn,Gauntlett:2009bh,Horowitz:2009ij}. It was also shown that adding higher-order terms to the scalar potential, as dictated by M-theory, can cure this problem. We expect the same to be the case here.

Furthermore we demonstrated that the holographic superfluid admits an entropy current, associated with its normal component, with positive definite divergence. Fundamentally, in the AdS/CFT setup, this of course follows from the second law of black hole thermodynamics and thus there should be a general proof to all orders in the gradient expansion. Such an an entropy current was developed in full generality for the gravity subsector in \cite{Bhattacharyya:2008xc} and it would be interesting to see if a similar analysis can be made for our solutions. We have reason to be optimistic about this, since the entropy of the field theory at zeroth order is simply given by the classical entropy associated with the black-hole horizon.

Finally, the analysis of this paper focuses on bulk solutions of $4d$ gravity, corresponding to thin films of superfluids. It would be interesting to repeat the numerical analysis in one higher dimension corresponding to relativistic $3+1$ dimensional superfluids which are also relevant for certain parts of the QCD phase diagram\footnote{Subject to the usual caveats of applying AdS/CFT techniques to QCD.}. Many of our analytical results are valid in any dimension and the remaining ones can be straightforwardly extended.

\subsection*{Acknowledgements}
\noindent
It is a pleasure to thank Evgeny Buchbinder, Jerome Gauntlett, Chris Herzog, James Lucietti, Andy O'Bannon, Toby Wiseman and Dam Son for useful discussions and correspondence.  We would also like to thank Amos Yarom for correspondence concerning the vanishing of the additional contributions to the field-theory entropy. J.S. thanks the EFI at the University of Chicago and the MCTP at the University of Michigan for hospitality at the beginning and finishing stages of this project. We thank Toby Wiseman for collaboration at an initial stage.
J.S. is supported by EPSRC and a JRF from Trinity College Cambridge, B.W. is supported by an STFC studentship.
%
\appendix
%

\section{The bulk equations}\label{app:bulkequations}
Here we reproduce the bulk equations governing the holographic superfluid. The equations for $f(r)$ and $h(r)$ are first-order ODEs, the rest are second order.
\vskip1cm

\underline{{\it$f$ equation:}}
\begin{footnotesize}
\bea
\lefteqn{-r^5 f \left(r \mathcal{B}'+4 \mathcal{B}-4\right)\; f'=}\nonumber\\
&& 4 r^5 f^2 \mathcal{B}'+4 r^4 h^2 \xi^2 \left(-\mathcal{B} f+\mathcal{C}^2+f\right)-12 r^4
   \left(f-h^2\right) \left(-\mathcal{B} f+\mathcal{C}^2+f\right)\nonumber\\
&&+2 r^6 f \xi '^2
   \left(-\mathcal{B} f+\mathcal{C}^2+f\right)+r^4 (\mathcal{B}-1) f \phi '^2-2 q^2 r^2
   (\mathcal{B}-1) h^2 \xi^2 \phi^2\nonumber\\
&&+r^6 (-f) \mathcal{C}'^2-8 r^5 \mathcal{C} f
   \mathcal{C}'-2 r^4 \mathcal{C} f \phi ' \Phi '\nonumber\\
&&+4 q^2 r^2 \mathcal{C} h^2 \xi^2 \phi 
   \Phi -2 q^2 r^2 f h^2 \xi ^2 \Phi ^2+r^4 f^2 \Phi '^2\nonumber\\
\eea
\end{footnotesize}
\underline{{\it$h$ equation:}}
\begin{footnotesize}
\bea
\lefteqn{4 r^3 f^2 \left(r \mathcal{B}'+4 \mathcal{B}-4\right)  \left(-\mathcal{B}
   f+\mathcal{C}^2+f\right)\; h'=}\nonumber\\
&& \mathcal{C}' \left(4 r^4 \mathcal{C} f^2 h \mathcal{B}'+16 r^3 \mathcal{C} f h
   \left(-\mathcal{B} f+\mathcal{C}^2+f\right)\right)\nonumber\\
&&+4 r^3 f h \mathcal{B}'\left(f-3
   h^2\right) \left(-\mathcal{B} f+\mathcal{C}^2+f\right)\nonumber\\
&&+\xi^2 \left(-4 r^3 f h^3
   \mathcal{B}' \left(-\mathcal{B} f+\mathcal{C}^2+f\right)-8 r^2 \mathcal{C}^2 h^3
   \left(-\mathcal{B} f+\mathcal{C}^2+f\right)\right)\nonumber\\
&&+\phi '\Phi ' \left(2 r^3 \mathcal{C}
   f^2 h \mathcal{B}'+4 r^2 \mathcal{C}^3 f(r) h\right)\nonumber\\
&&+\phi '^2 \left(r^3 (-(\mathcal{B}-1))
   f^2 h \mathcal{B}'-2 r^2 (\mathcal{B}-1) \mathcal{C}^2 f h\right)\nonumber\\
&&+\Phi '^2 \left(-r^3
   f^3 h \mathcal{B}'-2 r^2 \mathcal{C}^2 f^2 h\right)-2 r^4 f^3 h \mathcal{B}'^2\nonumber\\
&&+2 r^4
   f h \mathcal{C}'^2 \left(\mathcal{C}^2-2 (\mathcal{B}-1) f\right)-8 q^2 \mathcal{C}(r) h^3 \xi^2 \phi \Phi\left(\mathcal{C}^2-2 (\mathcal{B}-1) f\right)\nonumber\\
&&-4 q^2 (\mathcal{B}-1) h^3
   \xi^2 \phi^2 \left(2 (\mathcal{B}-1) f-\mathcal{C}^2\right)+4 q^2 f h^3 \xi^2 \Phi^2 \left(\mathcal{C}^2-2 (\mathcal{B}-1) f\right)\nonumber\\
&&-4 r^4 f h \xi '^2 \left(-3
   (\mathcal{B}-1) \mathcal{C}^2 f+2 (\mathcal{B}-1)^2 f^2+\mathcal{C}^4\right)\nonumber\\
&&+24 r^2 \mathcal{C}(r)^2
   h \left(f-h^2\right) \left(-\mathcal{B} f+\mathcal{C}^2+f\right)\nonumber\\
\eea

\end{footnotesize}
\underline{{\it${\cal B}$ equation:}}
\begin{footnotesize}
\bea
\lefteqn{4 r^8 f \left(-\mathcal{B} f+\mathcal{C}^2+f\right)\;\mathcal{B}''=}\nonumber\\
&& 8 r^8 \mathcal{C} f \mathcal{B}' \mathcal{C}'-4 r^7 h^2 \xi^2 \mathcal{B}' \left(-\mathcal{B}
   f+\mathcal{C}^2+f\right)\nonumber\\
&&-4 r^7 \mathcal{B}'\left(f+3 h^2\right) \left(-\mathcal{B}
   f+\mathcal{C}^2+f\right)+2 r^7 \mathcal{C} f \mathcal{B}' \phi ' \Phi '\nonumber\\
&&+\Phi '^2
   \left(4 r^6 f \left(-\mathcal{B} f+\mathcal{C}^2+f\right)-r^7 f^2 \mathcal{B}'\right)-4 r^8
   f^2 \mathcal{B}'^2\nonumber\\
&&-r^7 (\mathcal{B}-1) f \mathcal{B}' \phi '^2-4 r^8 (\mathcal{B}-1) f
   \mathcal{C}'^2\nonumber\\
&&+8 q^2 r^4 h^2 \xi (r)^2 \Phi^2 \left(-\mathcal{B} f+\mathcal{C}^2+f\right)\nonumber\\
\eea

\underline{{\it$\xi$ equation:}}
\begin{footnotesize}

\bea
\lefteqn{4 r^4 f \left(-\mathcal{B} f+\mathcal{C}^2+f\right)\;  \xi ''=}\nonumber\\
&& -4 r^3 h^2 \xi^2 \xi ' \left(-\mathcal{B} f+\mathcal{C}^2+f\right)-4 r^3 \left(f+3
   h^2\right) \xi ' \left(-\mathcal{B} f+\mathcal{C}^2+f\right)\nonumber\\
&&-8 r^2 h^2 \xi 
   \left(-\mathcal{B} f+\mathcal{C}^2+f\right)+r^3 (-(\mathcal{B}-1)) f(r) \xi ' \phi '^2+4 q^2
   (\mathcal{B}-1) h^2 \xi  \phi^2\nonumber\\
&&+2 r^3 \mathcal{C} f \xi ' \phi ' \Phi '-8 q^2
   \mathcal{C} h^2 \xi  \phi \Phi +4 q^2 f h^2 \xi \Phi^2-r^3 f^2 \xi ' \Phi
   '^2\nonumber\\
\eea
\end{footnotesize}

\underline{{\it$\Phi$ equation:}}
\begin{footnotesize}
\bea
\lefteqn{4 r^2 f  \left(-\mathcal{B} f+\mathcal{C}^2+f\right)\;\Phi ''=}\nonumber\\
&&\mathcal{B}' \left(4 r^2 \mathcal{C} f\phi '-4 r^2 f^2 \Phi '\right)+\mathcal{C}' \left(4 r^2
   \mathcal{C} f \Phi '-4 r^2 (\mathcal{B}-1) f\phi '\right)\nonumber\\
&&+8 q^2 h^2 \xi^2 \Phi
   \left(-\mathcal{B} f+\mathcal{C}^2+f\right)-4 r h^2 \xi^2 \Phi ' \left(-\mathcal{B}
   f+\mathcal{C}^2+f\right)\nonumber\\
&&+\Phi ' \left(4 r \left(f-3 h^2\right) \left(-\mathcal{B}
   f+\mathcal{C}^2+f\right)-r (\mathcal{B}-1) f \phi '^2\right)+2 r \mathcal{C} f \phi '
   \Phi '^2\nonumber\\
&&-r f^2 \Phi '^3\nonumber\\
\eea

\end{footnotesize}

\end{footnotesize}
\underline{{\it${\cal C}$ equation:}}
\begin{footnotesize}
\bea
\lefteqn{4 r^4 f \left(r \mathcal{B}'+4 \mathcal{B}-4\right) \left(-\mathcal{B}
   f+\mathcal{C}^2+f\right)\;\mathcal{C}'' =}\nonumber\\
&& \xi ^2 \phi  \Phi  \Big(-8 q^2 r h^2 \mathcal{B}' \left((\mathcal{B}-1)
   f+\mathcal{C}^2\right)+16 q^2 r (\mathcal{B}-1) \mathcal{C} h^2 \mathcal{C}'\nonumber\\
&& +32 q^2
   (\mathcal{B}-1) h^2 \left(-\mathcal{B} f+\mathcal{C}^2+f\right)\Big)\nonumber\\
&&+\xi ^2 \Phi ^2
   \left(8 q^2 r \mathcal{C} f h^2 \mathcal{B}'-8 q^2 r (\mathcal{B}-1) f h^2
   \mathcal{C}'\right)\nonumber\\
&&+\xi ^2 \Big(-4 r^4 h^2 \mathcal{B}' \mathcal{C}' \left(-\mathcal{B}
   f+\mathcal{C}^2+f\right)-16 r^3 \mathcal{C} h^2 \mathcal{B}' \left(-\mathcal{B}
   f+\mathcal{C}^2+f\right)\Big)\nonumber\\
&&+\mathcal{C}' \Big(4 r^4 \mathcal{B}' \left((\mathcal{B}-1) f
   \left(f+3 h^2\right)+\mathcal{C}^2 \left(7 f-3 h^2\right)\right)\nonumber\\
&&-4 r^5 f^2
   \mathcal{B}'^2-64 r^3 (\mathcal{B}-1) f \left(-\mathcal{B} f+\mathcal{C}^2+f\right)\Big)\nonumber\\
&&+\xi
   '^2 \left(8 r^5 (\mathcal{B}-1) f \mathcal{C}' \left(-\mathcal{B} f+\mathcal{C}^2+f\right)-8
   r^5 \mathcal{C} f \mathcal{B}' \left(-\mathcal{B}
   f+\mathcal{C}^2+f\right)\right)\nonumber\\
&&+\mathcal{C}'^2 \left(8 r^5 \mathcal{C} f \mathcal{B}'-16 r^4
   (\mathcal{B}-1) \mathcal{C} f\right)\nonumber\\
&&+\phi '^2 \left(r^4 (-(\mathcal{B}-1)) f \mathcal{B}'
   \mathcal{C}'-4 r^3 (\mathcal{B}-1) \mathcal{C} f \mathcal{B}'\right)\nonumber\\
&&+\Phi '^2 \left(r^4
   \left(-f^2\right) \mathcal{B}' \mathcal{C}'-4 r^3 \mathcal{C} f^2 \mathcal{B}'\right)\nonumber\\
&&+\phi '
   \Phi ' \Big(2 r^4 \mathcal{C} f \mathcal{B}' \mathcal{C}'+4 r^3 f \mathcal{B}'
   \left(-\mathcal{B} f+3 \mathcal{C}^2+f\right)\nonumber\\
&&+16 r^2 (\mathcal{B}-1) f \left(-\mathcal{B}
   f+\mathcal{C}^2+f\right)\Big)\nonumber\\
&&+\xi ^2 \phi ^2 \left(8 q^2 r (\mathcal{B}-1) \mathcal{C}
   h^2 \mathcal{B}'-8 q^2 r (\mathcal{B}-1)^2 h^2 \mathcal{C}'\right)\nonumber\\
&&+48 r^3 \mathcal{C}
   \mathcal{B}' \left(f-h^2\right) \left(-\mathcal{B} f+\mathcal{C}^2+f\right)-16 r^4
   \mathcal{C}f^2 \mathcal{B}'^2-4 r^5 (\mathcal{B}-1) f\mathcal{C}'^3\nonumber\\
\eea

\end{footnotesize}
\underline{{\it$\phi$ equation:}}
\begin{footnotesize}
\bea
\lefteqn{4 r^3 f \left(r \mathcal{B}'+4 \mathcal{B}-4\right) \left(-\mathcal{B}
   f+\mathcal{C}^2+f\right)\; \phi '' =}\nonumber\\
&& \mathcal{C}' \Big(\mathcal{B}' \left(4 r^4 \mathcal{C} f \phi '-4 r^4 f^2 \Phi '\right)-16 r^3
   (\mathcal{B}-1) \mathcal{C} f \phi '\nonumber\\
&&+16 r^3 f \Phi ' \left(-\mathcal{B} f+2
   \mathcal{C}^2+f\right)\Big)\nonumber\\
&&+\phi  \Big(\xi ^2 \left(8 q^2 r^2 h^2 \mathcal{B}'
   \left(-\mathcal{B} f+\mathcal{C}^2+f\right)+32 q^2 r (\mathcal{B}-1) h^2 \left(-\mathcal{B}
   f+\mathcal{C}^2+f\right)\right)\nonumber\\
&&+\xi ^2 \Phi  \left(16 q^2 (\mathcal{B}-1) \mathcal{C} h^2
   \phi '-16 q^2 \mathcal{C}^2 h^2 \Phi '\right)\Big)\nonumber\\
&&+\mathcal{B}' \Big(4 r^3 \phi ' \left(3
   (\mathcal{B}-1) f \left(f+h^2\right)+\mathcal{C}^2 \left(f-3 h^2\right)\right)\nonumber\\
&&+r^3
   (-(\mathcal{B}-1)) f \phi '^3+\Phi ' \left(2 r^3 \mathcal{C} f \phi '^2-16 r^3 \mathcal{C}
   f^2\right)-r^3 f^2 \phi ' \Phi '^2\Big)\nonumber\\
&&+\xi ^2 \left(-4 r^3 h^2 \mathcal{B}' \phi '
   \left(-\mathcal{B} f+\mathcal{C}^2+f\right)-16 r^2 \mathcal{C} h^2 \Phi ' \left(-\mathcal{B}
   f+\mathcal{C}^2+f\right)\right)\nonumber\\
&&+\mathcal{C}'^2 \left(4 r^4 \mathcal{C} f \Phi '-4 r^4
   (\mathcal{B}-1) f \phi '\right)\nonumber\\
&&+\xi ^2 \Phi ^2 \left(8 q^2 \mathcal{C} f h^2 \Phi '-8
   q^2 (\mathcal{B}-1) f h^2 \phi '\right)\nonumber\\
&&+\Phi ' \left(48 r^2 \mathcal{C}
   \left(f-h^2\right) \left(-\mathcal{B} f+\mathcal{C}^2+f\right)-4 r^2 (\mathcal{B}-1)
   \mathcal{C} f \phi '^2\right)\nonumber\\
&&+\xi '^2 \left(-8 r^4 (\mathcal{B}-1) f \phi '
   \left((\mathcal{B}-1) f-\mathcal{C}^2\right)-8 r^4 \mathcal{C} f \Phi ' \left(-\mathcal{B}
   f+\mathcal{C}^2+f\right)\right)\nonumber\\
&&-32 r^2 (\mathcal{B}-1) f \phi ' \left(-\mathcal{B}
   f+\mathcal{C}^2+f\right)\nonumber\\
&&+\xi ^2 \phi ^2 \left(8 q^2 (\mathcal{B}-1) \mathcal{C} h^2 \Phi
   '-8 q^2 (\mathcal{B}-1)^2 h^2 \phi '\right)+8 r^2 \mathcal{C}^2 f \phi ' \Phi '^2\nonumber\\
&&-4 r^2
   \mathcal{C} f^2 \Phi '^3\nonumber\\
\eea

\end{footnotesize}

\bibliography{mybib}{}

\begin{thebibliography}{10}

\bibitem{Hartnoll:2008vx}
Sean~A. Hartnoll, Christopher~P. Herzog, and Gary~T. Horowitz.
\newblock {\it Building a Holographic Superconductor}.
\newblock {\em Phys. Rev. Lett.}, 101:031601, 2008.

\bibitem{Hartnoll:2008kx}
Sean~A. Hartnoll, Christopher~P. Herzog, and Gary~T. Horowitz.
\newblock {\it Holographic Superconductors}.
\newblock {\em JHEP}, 12:015, 2008.

\bibitem{Gubser:2008px}
Steven~S. Gubser.
\newblock {\it Breaking an Abelian Gauge Symmetry Near a Black Hole Horizon}.
\newblock {\em Phys. Rev.}, D78:065034, 2008.

\bibitem{Maldacena:1997re}
Juan~Martin Maldacena.
\newblock {\it The Large $N$ Limit of Superconformal Field Theories and
  Supergravity}.
\newblock {\em Adv. Theor. Math. Phys.}, 2:231--252, 1998.

\bibitem{Witten:1998zw}
Edward Witten.
\newblock {\it Anti-de~Sitter Space, Thermal Phase Transition, and Confinement
  in Gauge Theories}.
\newblock {\em Adv. Theor. Math. Phys.}, 2:505--532, 1998.

\bibitem{Policastro:2001yc}
G.~Policastro, D.~T. Son, and A.~O. Starinets.
\newblock {\it The Shear Viscosity of Strongly Coupled ${\mathcal{N}}\!=4$
  Supersymmetric Yang-Mills Plasma}.
\newblock {\em Phys. Rev. Lett.}, 87:081601, 2001.

\bibitem{Bhattacharyya:2008jc}
Sayantani Bhattacharyya, Veronika~E Hubeny, Shiraz Minwalla, and Mukund
  Rangamani.
\newblock {\it Nonlinear Fluid Dynamics from Gravity}.
\newblock {\em JHEP}, 02:045, 2008.

\bibitem{Banerjee:2008th}
Nabamita Banerjee et~al.
\newblock {\it Hydrodynamics from Charged Black Branes}.
\newblock 2008.

\bibitem{Erdmenger:2008rm}
Johanna Erdmenger, Michael Haack, Matthias Kaminski, and Amos Yarom.
\newblock {\it Fluid Dynamics of R-Charged Black Holes}.
\newblock {\em JHEP}, 01:055, 2009.

\bibitem{Hansen:2008tq}
James Hansen and Per Kraus.
\newblock {\it Nonlinear Magnetohydrodynamics from Gravity}.
\newblock {\em JHEP}, 04:048, 2009.

\bibitem{Landau:1941}
Lev~Davidovich Landau.
\newblock {\it Theory of the Superfluidity of Helium II}.
\newblock {\em Phys. Rev.}, 60:356, 1941.

\bibitem{Tisza:1947zz}
Laszlo Tisza.
\newblock {\it The Theory of Liquid Helium}.
\newblock {\em Phys. Rev.}, 72:838--854, 1947.

\bibitem{Chamblin:1999tk}
Andrew Chamblin, Roberto Emparan, Clifford~V. Johnson, and Robert~C. Myers.
\newblock {\it Charged AdS Black Holes and Catastrophic Holography}.
\newblock {\em Phys. Rev.}, D60:064018, 1999.

\bibitem{Herzog:2008he}
C.~P. Herzog, P.~K. Kovtun, and D.~T. Son.
\newblock {\it Holographic model of superfluidity}.
\newblock {\em Phys. Rev.}, D79:066002, 2009.

\bibitem{Basu:2008st}
Pallab Basu, Anindya Mukherjee, and Hsien-Hang Shieh.
\newblock {\it Supercurrent: Vector Hair for an AdS Black Hole}.
\newblock {\em Phys. Rev.}, D79:045010, 2009.

\bibitem{Herzog:2009md}
Christopher~P. Herzog and Amos Yarom.
\newblock {\it Sound Modes in Holographic Superfluids}.
\newblock {\em Phys. Rev.}, D80:106002, 2009.

\bibitem{Karch:2008fa}
A.~Karch, D.~T. Son, and A.~O. Starinets.
\newblock {\it Zero Sound from Holography}.
\newblock 2008.

\bibitem{Yarom:2009uq}
Amos Yarom.
\newblock {\it Fourth Sound of Holographic Superfluids}.
\newblock {\em JHEP}, 07:070, 2009.

\bibitem{Amado:2009ts}
Irene Amado, Matthias Kaminski, and Karl Landsteiner.
\newblock {\it Hydrodynamics of Holographic Superconductors}.
\newblock {\em JHEP}, 05:021, 2009.

\bibitem{Herzog:2009ci}
Christopher~P. Herzog and Silviu~S. Pufu.
\newblock {\it The Second Sound of $SU(2)$}.
\newblock {\em JHEP}, 04:126, 2009.

\bibitem{Arean:2010xd}
Daniel Arean, Matteo Bertolini, Jarah Evslin, and Tomas Prochazka.
\newblock {\it On Holographic Superconductors with DC Current}.
\newblock 2010.

\bibitem{Horowitz:2008bn}
Gary~T. Horowitz and Matthew~M. Roberts.
\newblock {\it Holographic Superconductors with Various Condensates}.
\newblock {\em Phys. Rev.}, D78:126008, 2008.

\bibitem{Franco:2009yz}
Sebastian Franco, Antonio Garc{\'\i a-}Garcia, and Diego Rodriguez-G\'omez.
\newblock {\it A General Class of Holographic Superconductors}.
\newblock 2009.

\bibitem{Gauntlett:2009dn}
Jerome~P. Gauntlett, Julian Sonner, and Toby Wiseman.
\newblock {\it Holographic superconductivity in M-Theory}.
\newblock {\em Phys. Rev. Lett.}, 103:151601, 2009.

\bibitem{Gauntlett:2009bh}
Jerome Gauntlett, Julian Sonner, and Toby Wiseman.
\newblock {\it Quantum Criticality and Holographic Superconductors in M-
  theory}.
\newblock {\em JHEP}, 2009.

\bibitem{Gubser:2009qm}
Steven~S. Gubser, Christopher~P. Herzog, Silviu~S. Pufu, and Tiberiu Tesileanu.
\newblock {\it Superconductors from Superstrings}.
\newblock {\em Phys. Rev. Lett.}, 103:141601, 2009.

\bibitem{Balasubramanian:1999re}
Vijay Balasubramanian and Per Kraus.
\newblock {\it A stress tensor for anti-de Sitter gravity}.
\newblock {\em Commun. Math. Phys.}, 208:413--428, 1999.

\bibitem{Gibbons:1976ue}
G.~W. Gibbons and S.~W. Hawking.
\newblock {\it Action Integrals and Partition Functions in Quantum Gravity}.
\newblock {\em Phys. Rev.}, D15:2752--2756, 1977.

\bibitem{Son:2000ht}
D.~T. Son.
\newblock {\it Hydrodynamics of relativisic systems with broken continuous
  symmetries}.
\newblock {\em Int. J. Mod. Phys.}, A16S1C:1284--1286, 2001.

\bibitem{Son:2000ht2}
D.~T. Son.
\newblock {\it Relativistic hydrodynamics of systems with spontaneous symmetry
  breaking: the Poisson bracket approach}.
\newblock {\em unpublished}, 2001.

\bibitem{LLvol6}
Evgeny~Mikhailovich Lifshitz and Lev~Davidovich Landau.
\newblock {\em Course of Theoretical Physics}, volume~6.
\newblock 1959.

\bibitem{Bhattacharyya:2008xc}
Sayantani Bhattacharyya et~al.
\newblock {\it Local Fluid Dynamical Entropy from Gravity}.
\newblock {\em JHEP}, 06:055, 2008.

\bibitem{Sonner:2009fk}
Julian Sonner.
\newblock {\it A Rotating Holographic Superconductor}.
\newblock {\em Phys. Rev.}, D80:084031, 2009.

\bibitem{Horowitz:2009ij}
Gary~T. Horowitz and Matthew~M. Roberts.
\newblock {\it Zero Temperature Limit of Holographic Superconductors}.
\newblock {\em JHEP}, 11:015, 2009.

\end{thebibliography}
\bibliographystyle{unsrt}

\end{document}